\documentclass[a4paper,11pt]{article}
\usepackage{graphicx} 
\usepackage[utf8]{inputenc}
\usepackage{xcolor}
\usepackage{authblk}
\usepackage{hyperref}
\usepackage{comment}
\usepackage{euscript}
\DeclareMathAlphabet{\mathpzc}{T1}{pzc}{m}{it}
\hypersetup{pdfborder=0 0 0}

\usepackage[backend=biber, style = phys]{biblatex}
\usepackage{soul}
\usepackage{mathtools}
\usepackage{graphicx}
\usepackage{booktabs}
\usepackage{subcaption}
\usepackage{amsmath}
\usepackage{amssymb}
\usepackage{xfrac}  
\usepackage[utf8]{inputenc}

\graphicspath{ {./images/} }
\definecolor{raspberryrose}{rgb}{0.7, 0.27, 0.42}
\definecolor{skobeloff}{rgb}{0.0, 0.48, 0.45}

\title{A minimal physical model of cochlear mechanics: Insights into active nonlinear feedback}

\author{Vidyunmathi N A\thanks{\texttt{p20220022@goa.bits-pilani.ac.in ; vidyu.bhat@gmail.com}} }
\author{Toby Joseph\thanks{\texttt{toby@goa.bits-pilani.ac.in}}}

\affil{Department of Physics, BITS Pilani K K Birla Goa Campus}
\date{}

\begin{document}

\maketitle{}
The remarkable sensitivity, compressive nonlinearity, and frequency selectivity of the mammalian 
cochlea arise from an active process that amplifies the passive mechanical response of the basilar 
membrane. Outer hair cells are widely regarded as the primary effectors of this active process in 
the basal regions of the cochlea. This active contribution progressively decreases with increasing 
stimulus level. Motivated by these observations, a minimal cochlear model is investigated in which 
each location is represented by a forced damped oscillator with an exponential displacement-dependent 
active feedback. The oscillators are coupled through the cochlear fluid and also by elastic and 
dissipative longitudinal interactions to form a one-dimensional distributed model. The model 
reproduces key features of cochlear mechanics, including level-dependent amplification, 
compressive nonlinearity, frequency selectivity, traveling-wave propagation, and phase 
accumulation. Comparison with a simplified cubic nonlinear model has been carried out. 
Even though the cubic nonlinearity describes the response for weak stimulus, at higher 
stimulus levels it deviates from the exponential model and fails to reproduce the gradual 
transition towards predominantly passive behavior. Longitudinal coupling broadens the frequency 
response, modifies the traveling wave profile, and increases phase accumulation. The proposed 
model provides a simple physical framework for understanding how local active processes and 
longitudinal mechanical interactions together shape the nonlinear response of the cochlea.

\section*{Introduction}
The auditory system is of considerable scientific and clinical interest owing to its remarkable 
performance and importance in everyday communication. The extraordinary sensitivity, temporal 
precision, frequency resolution, and dynamic range of hearing reflect a highly optimized 
biological signal-processing system. Understanding the physical mechanisms that underlie these 
capabilities has been a longstanding objective in auditory research.  
\par
The primary mechanical processing of the sound takes place in the cochlea, where the basilar 
membrane responds to acoustic stimuli and generates a traveling wave along the longitudinal 
direction. Owing to the tonotopic organization of the cochlea, the traveling waves of different 
frequencies peak at different locations \cite{reichenbach2014physics,OLSON2020392}.
 The pioneering experiments of Von Békésy established the traveling wave as a fundamental feature 
 of cochlear mechanics, providing the experimental basis for subsequent theoretical descriptions 
 of cochlear wave propagation \cite{10.1121/1.1906503, OLSON201231}.
 \par
The first mathematical descriptions of cochlear mechanics demonstrated that a traveling wave 
arises from the interaction between the longitudinally varying mechanical properties of the 
cochlear partition and the surrounding fluid. Passive descriptions related the spatial evolution 
of traveling wave amplitude, wavelength, and phase to the local mechanical response of the 
cochlear partition and its coupling through the fluid \cite{zweig1976basilar}. Subsequent two 
and three-dimensional hydrodynamical models provided more detailed treatments of fluid-partition 
interaction and were evaluated against experimental measurements and physical cochlear models    
 \cite{10.1121/1.381272,10.1121/1.382570}. However, passive mechanics by itself cannot account 
 for the remarkable sensitivity and the sharp frequency selectivity of the living cochlea. 
 Responses measured in the living cochlea are strongly amplified, sharply tuned, and grow compressively 
 over a wide range of sound intensities. These observations indicate the presence of an active 
 mechanism in the cochlea, which enhances weak responses and compensates dissipative 
 losses \cite{ashmore1989mechanics, OLSON2020392}.

 The recognition of the active mechanics \cite{rhode1971observations, kemp1978stimulated} motivated the development of models incorporating active feedback into the passive system, enhancing the sensitivity and frequency selectivity of the response.  Early linear descriptions of the active process reproduced the sharp tuning and amplification observed at low stimulus levels \cite{neely1985mathematical,nobili1993Linear}. Subsequent nonlinear extensions incorporated saturation of the active feedback and reproduced the compressive growth of basilar-membrane responses \cite{10.1121/1.415412}.
 \par
 Models based on Van der Pol and Rayleigh-type oscillators constitute another important class of cochlear models \cite{diependaal1983department,duifhuis1986modelling}. 
 Beginning with these formulations, nonlinear active feedback was shown to reproduce amplification, compressive growth, and other qualitative features of cochlear mechanics.
 Subsequent developments explored different forms of nonlinear feedback, extending the range of cochlear responses captured within this framework\cite{van1990generalized,duifhuis2012cochlear}.
 A different yet closely related perspective emerged from the study of nonlinear dynamical systems, where the characteristic properties of the active process were examined in terms of a critical oscillator. A critical oscillator near a Hopf bifurcation exhibits strong amplification of weak inputs, compressive response growth, and level-dependent frequency selectivity. Several essential features of active hearing arise from a common mechanism in this dynamical framework \cite{PhysRevLett.84.5232, doi:10.1073/pnas.97.7.3183, hudspeth2010critique,kern2003essential}. Different approaches have coupled tonotopically distributed nonlinear oscillators through cochlear hydrodynamics or combined them with reduced descriptions of directional wave propagation, demonstrating how local nonlinear activity and wave propagation together shape amplification, compression, and frequency tuning \cite{PhysRevLett.90.158101}. More recently, the collective behavior of distributed critical oscillators has been examined in the presence of longitudinal mechanical coupling and energy pumping, emphasizing the distinct roles of local activity and spatial interactions in shaping cochlear tuning\cite{hulst2025marrying,kern2003essential}.
 \par
 The present work adopts a minimal modeling approach to investigate which characteristic response properties of the active cochlea can emerge from a spatially distributed system of nonlinear active oscillators. The analysis begins with a single forced and damped oscillator incorporating a nonlinear amplitude-dependent active feedback term. The local elements are then distributed along a tonotopic axis and coupled through the pressure field. Additional longitudinal coupling is subsequently introduced to examine how spatial interaction along the oscillator chain modifies the response. This formulation retains a small set of physically interpretable ingredients - local resonance, active compensation of damping, amplitude-dependent saturation, fluid-mediated interaction, and longitudinal coupling. 
 \par
Traveling wave propagation, phase accumulation, level-dependent amplification, compressive growth, and frequency selectivity are investigated from the responses of the oscillator chain. The contribution of higher-order nonlinearities is assessed by comparing the complete saturating active term with its cubic approximation. The influence of longitudinal coupling on amplification, frequency selectivity, and phase accumulation is then investigated. The model responses are compared qualitatively with experimental measurements of basilar-membrane motion in the chinchilla cochlea \cite{ruggero1997basilar}, with parameters chosen to approximate the corresponding tonotopic and mechanical properties.
\section*{The minimal model}
In the present model, the basilar membrane - organ of Corti system (BM-OC) is a one-dimensional inhomogeneous chain of oscillators coupled through the cochlear fluid. 
\par
\subsubsection*{Local single oscillator}

Each oscillator represents a local cross-section of the BM-OC system and is described by a forced under-damped oscillator with a displacement-dependent nonlinear behavior.
\begin{equation}
    \Ddot{\eta}(t) = -\gamma  \Dot{\eta}(t) +  \beta e^{-\left(\frac{\eta(t)}{\eta_{\rm th}}\right)^2} \Dot{\eta}(t) - {{\omega}_0}^2 \eta(t) + F_0 \sin{{\omega}t}.
    \label{eq1}
\end{equation}
Here, $\eta(t)$ denotes the displacement of the oscillator, $\omega_{0}$ is its natural frequency, and $\gamma$ is the positive damping coefficient. The term 
$ \beta e^{-\left(\frac{\eta(t)}{\eta_{\rm th}}\right)^2} \Dot{\eta}(t)$ represents the active process, where $\beta$ is the strength of active amplification. The active contribution opposes the dissipative damping of the oscillator. Its contribution is greatest for small displacements, and progressively decreases as the displacement increases. Consequently, weak responses experience the greatest active amplification. $\eta_{\rm th}$ sets the displacement scale beyond which active contribution weakens.  The oscillator is driven by an external sinusoidal force of amplitude $F_0$ and frequency $\omega$.
\par
This nonlinear oscillator captures key features associated with the active process of the inner ear.  This single oscillator system undergoes supercritical Hopf bifurcation at $ \beta = \gamma $. For $ \beta < \gamma $, the equilibrium state is stable and the system undergoes damped motion. For $ \beta > \gamma$, the equilibrium becomes unstable, and the system exhibits self-sustained oscillations. The vicinity of the bifurcation is associated with enhanced sensitivity and compressive nonlinear behavior, both of which are characteristic features of the active cochlear mechanics. 
\par
\subsubsection*{Coupled oscillator Model}
To model the BM-OC system, an inhomogeneous chain of local oscillators is considered along the cochlea. The displacements of the oscillators are described by the field $\eta(x,t)$, where $x$ denotes the longitudinal position along the cochlea. The external forcing is introduced via the fluid pressure difference across the partition, $P(x,t)$. This, in effect, couples the oscillators through the cochlear fluid.
The equation governing the dynamics of the distributed system is,
\begin{align}
\Ddot{\eta}(x,t) =&
-\omega_0(x)^2 \eta(x,t)
-\gamma \Dot{\eta}(x,t)
+\beta e^{-\left(\frac{\eta(x,t)}{\eta_{\rm th}}\right)^2}
\Dot{\eta}(x,t)
\nonumber\\[2mm]
&
+\frac{A}{m(x)}P(x,t)
.
\label{eq2}
\end{align}
For weak response amplitudes, the active feedback reduces to $\beta {\left[1-\left(\frac{\eta(x,t)}{\eta_{\rm th}}\right)^2\right]}$, similar to the cubic nonlinear feedback employed in many Van der Pol and Rayleigh-type cochlear oscillator models. The quantities $\beta$, $\gamma$, and $\eta_{\rm th}$ retain the same interpretation as in the 
single oscillator model. The pressure field $P(x,t)$ provides the forcing on the cochlear partition, 
while $A$ and $m(x)$ denote the effective area and mass associated with a local oscillator segment. 
The natural frequency of each oscillator is tonotopically decided, where $\omega_0(x) = \omega_b \exp(-\mu x)$. 
It decreases longitudinally along the length of the basilar membrane. $\omega_b$ is the natural frequency of 
the first oscillator at the base and $\mu$ is a characteristic length scale.
Assuming an incompressible and inviscid cochlear fluid and adapting one-dimensional hydrodynamic 
approximation, the pressure difference across the basilar membrane satisfies \cite{PhysRevLett.90.158101}, 
\begin{equation}
                    \frac{2 \rho}{H}\Ddot{\eta}(x,t) = \frac{\partial^2 P(x,t)}{\partial x^2} \;,
                    \label{eq3}
\end{equation}
where $\rho$ is the fluid density and $H$ is height of the cochlear partition. 
 To determine the evolutions of pressure and displacement fields along the cochlea, Eqs. \ref{eq2} 
 and \ref{eq3} are solved simultaneously. Appropriate boundary conditions are imposed at the basal 
 and apical ends of the cochlear partition.  The fluid pressure drives the motion of the cochlear 
 partition through Eq. \ref{eq2}, while the displacement of the partition modifies the pressure 
 distribution through Eq. \ref{eq3}. This bidirectional coupling gives rise to traveling wave 
 propagation along the oscillator chain. 
\subsubsection*{Longitudinal Coupling}
Although fluid coupling alone generates traveling wave propagation, the resulting response remains strongly 
localized near the characteristic place. To model the direct mechanical interactions between the neighboring 
oscillators along the cochlear partition, elastic and dissipative longitudinal coupling are introduced. 
The corresponding coupling strengths are denoted by $\alpha$ and $\alpha'$ respectively. 
The elastic coupling promotes energy exchange between neighboring oscillators, broadens the spatial response, 
and modifies the frequency selectivity of the system. The dissipative coupling provides the damping associated 
with the relative motion of neighboring oscillators, thereby suppressing short-wavelength spatial oscillations. 
Incorporating these longitudinal interactions modifies the equation of motion,

\begin{align}
\Ddot{\eta}(x,t) =&
-\omega_0(x)^2 \eta(x,t)
-\gamma \Dot{\eta}(x,t)
+\beta e^{-\left(\frac{\eta(x,t)}{\eta_{\rm th}}\right)^2}
\Dot{\eta}(x,t)
\nonumber\\
&
+\frac{\alpha}{m(x)}
\frac{\partial^2 \eta(x,t)}{\partial x^2}
+\frac{\alpha'}{m(x)}
\frac{\partial^2 \Dot{\eta}(x,t)}{\partial x^2}
+\frac{A}{m(x)}P(x,t) \;.
\label{eq4}
\end{align}
\subsubsection*{Numerical Implementation}

The governing equations were solved using the Method of Lines. The BM-OC system was discretized 
into $N_x$ spatial points, distributed uniformly along the longitudinal direction, and the spatial 
derivatives were approximated using finite-difference operators. A harmonic pressure stimulus was applied 
at the basal boundary according to,  $P(0,t) = P_0\sin{\omega t}$, while the pressure difference at the 
apical boundary was maintained at zero, $P(L,t) = 0$. To account for the transfer function of the middle ear, an additional gain of 30 dB was applied to the stimulus pressure before it was used to drive the cochlear model\cite{ravicz2013inner,ravicz2013middle,slama2010middle}.
\par
The first oscillator at the base was fixed, $\eta(0,t) = 0$, whereas the remaining oscillators were evolved 
according to the governing equations. At each time step, the pressure field was obtained from the discretized 
fluid equation and subsequently used to evaluate the acceleration of the BM-OC system. The resulting system was 
integrated using MATLAB's \texttt{ode45} solver until a steady periodic response was established. Response amplitudes 
and phases were extracted from the final oscillation cycles after transient effects had decayed. 
The simulations were performed using physiological parameters representative of the chinchilla cochlea \cite{santi1979morphometry,robles2001mechanics,muller2010physiological,ver2024nonlinear,ruggero1997basilar}. 
The physiological and model parameters used in the simulations are listed in 
Table \ref{parameters_tab}.

\begin{table}[t!]
\centering
\caption{Model parameters}
\begin{tabular}{ll}

Parameter & Value  \\
\midrule

Basilar membrane Length, $L$ & 1.85 cm  \\
Mass of BM-OC segment, $m$ & $2 \times 10^{-7}$ cm \\
Area of the BM-OC segment, $A$  & $2 \times 10^{-5}$ \text{cm}$^{2}$ \\
Fluid chamber height, $H$ & 0.1 cm  \\
Fluid density, $\rho$ & 1 g \text{cm}$^{-3}$  \\
 Frequency length scale, $\mu$ & $3.1$ \text{cm}$^{-1}$ \\
Damping coefficient, $\gamma$ & $12000$ s$^{-1}$ \\
Active feedback strength, $\beta$ & $0.99$ $\gamma $  \\
Displacement threshold, $\eta_{th}$ & $8\times10^{-7}$ cm  \\
Elastic coupling coefficient, $\alpha$ & $3.0\times10^{-4}$ g \text{cm}$^{2}$ \text{s}$^{-2}$    \\
Dissipative coupling coefficient, $\alpha'$ & $1.2\times10^{-9}$ g  \text{cm}$^{2}$  \text{s}$^{-1}$   \\
Number of segments, $N_x$ & 2000  \\
\bottomrule
\end{tabular}
\label{parameters_tab}
\end{table}
 \section*{Results}
 \begin{figure}[h]
    \centering
    \includegraphics[width=\linewidth]{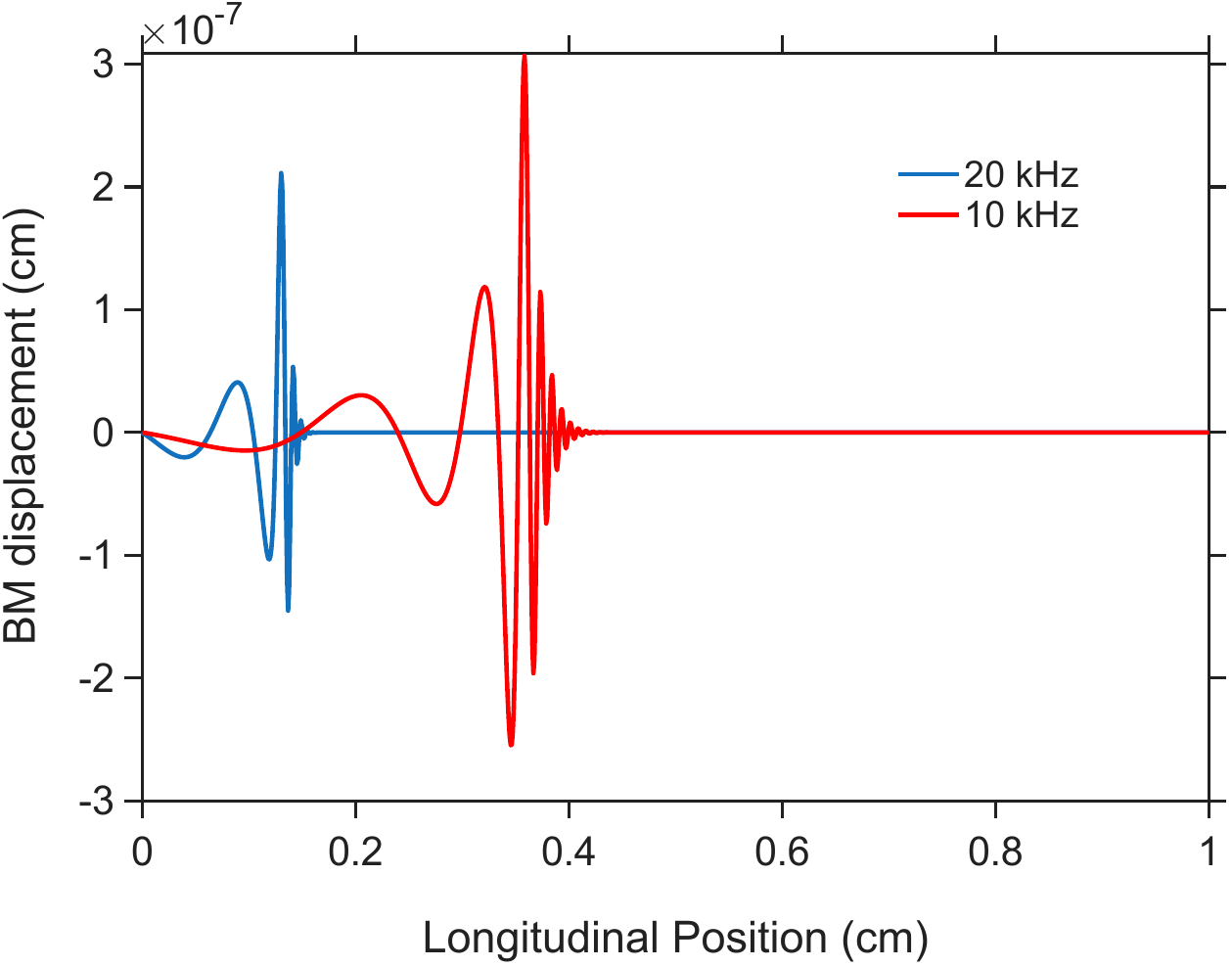}
    \caption{Spatial displacement profiles of the traveling wave for $20$ kHz and $10$ kHz stimuli at $50$ $\text{dB  SPL}$. 
    The figure illustrates the frequency selectivity of the model. For clarity, only the relevant portion of the basilar 
    membrane(BM) is shown.}
    \label{fig:TW}
\end{figure}

\subsubsection*{Traveling Wave Response}
 
Fig. \ref{fig:TW} shows the spatial displacement profiles for two stimulus frequencies. The location of the maximum 
response varies systematically with the stimulus frequency. The $20$ kHz stimulus reaches its peak close to the basal end 
of the cochlea, whereas the $10$ kHz stimulus propagates further before attaining its maximum response. This behavior 
is a consequence of the tonotopic organization of the cochlea, in which each longitudinal position is associated with 
a characteristic frequency (CF). In the present model, the passive tonotopic organization is prescribed through the local 
resonant frequency, $\sqrt{k(x)/m(x)}$. Mass is assumed to be constant longitudinally. The stiffness is assumed to decrease 
exponentially from the base towards the apex, producing a corresponding variation in the local passive resonant frequency. 
The traveling wave response emerges from the interaction of the spatially distributed oscillators with the cochlear fluid, 
and together with the longitudinal interaction between the neighboring oscillators. Consequently, for each stimulus frequency, 
the traveling wave reaches its maximum near the longitudinal location where the CF matches the applied 
stimulus frequency. Higher frequencies peak near the base, whereas lower frequencies peak towards the apex.

\subsubsection*{Compressive Nonlinearity}

\begin{figure}[h]
\centering
\includegraphics[width=\linewidth]{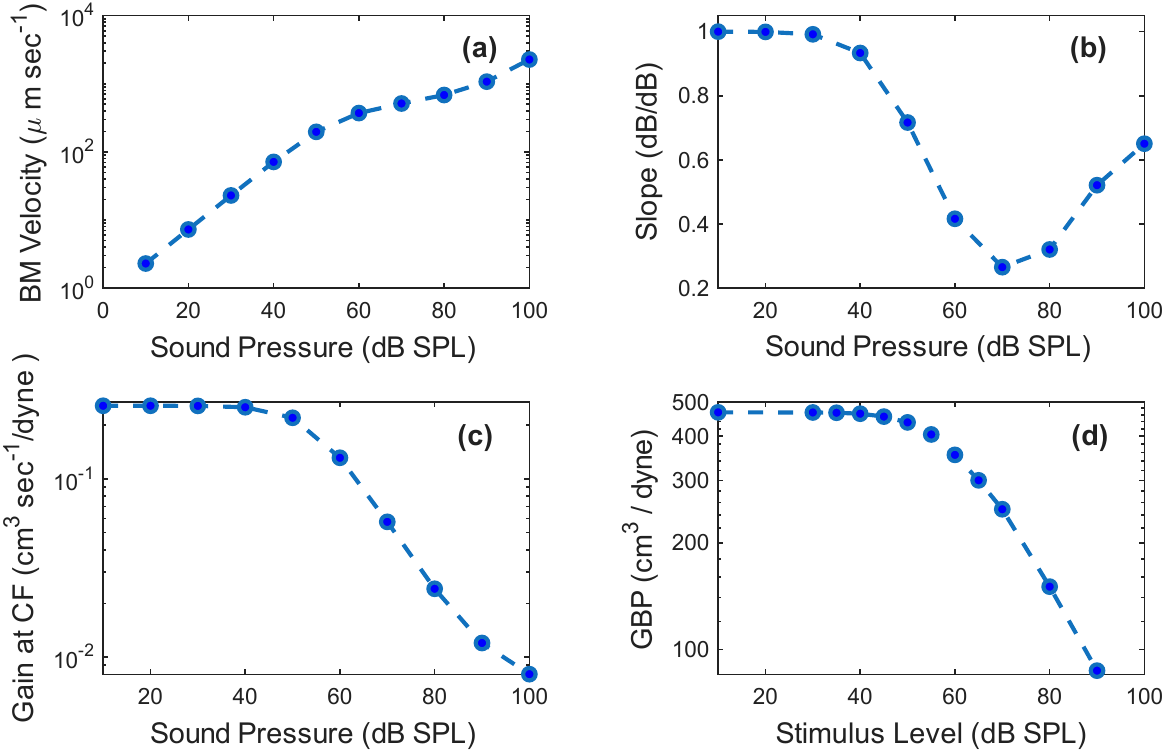}
\caption{
Model response illustrating level-dependent nonlinear compression at the location where CF $= 10$ kHz.
(a) Velocity-response function.
(b) Response slope (c) Gain  (d) Gain-Bandwidth product.}
\label{fig:compression_panel}
\end{figure}

Figure \ref{fig:compression_panel} summarizes the level-dependent velocity response at the location on the BM-OC 
corresponding to CF of $10$ kHz. The location of CF is defined as 
the position where there is maximal response for a stimulus at that frequency as the stimulus intensity goes to zero.
The velocity response, local slope, gain, and gain-bandwidth product together characterize the compressive nonlinearity produced 
by the active process. The four quantities provide complementary measures of the same underlying transition 
from linear amplification at low stimulus levels to compressive nonlinear behavior at higher stimulus levels.

Figure $\ref{fig:compression_panel}$(a) shows the velocity-intensity function of the basilar membrane at 
the CF location. The response grows approximately linearly up to about $30$ \text{dB SPL}. 
As the stimulus level increases, the corresponding growth becomes progressively slower, indicating the onset 
of compressive nonlinearity. The local rate of growth of the response is shown in Figure $\ref{fig:compression_panel}(b)$. At low stimulus levels, 
the slope remains close to unity. With the increase in stimulus level, the local slope decreases, 
reaching a minimum of $0.26$ $\text{dB/dB}$ at $70$ dB SPL before increasing again for higher stimulus levels. 
The same transition is reflected in the gain at the CF as shown in Fig. \ref{fig:compression_panel}(c).
Gain is defined as the ratio of the response amplitude to the applied pressure amplitude.
At low stimulus levels, the gain remains approximately constant. In the compressive regime, the gain decreases monotonically with 
increasing stimulus level as progressively larger responses receive less active amplification. 
Fig. \ref{fig:compression_panel}(d) shows the variation of the gain-bandwidth product with the stimulus levels. 
The gain-bandwidth product remains approximately constant at low stimulus levels and decreases once the system 
enters the compressive regime. The qualitative trend is consistent with experimental observations \cite{ruggero1997basilar,ver2024marrying}. 
The reduction in gain with increasing stimulus level is accompanied by a comparatively smaller change in bandwidth, leading to 
a decrease in the gain-bandwidth product.

\subsubsection*{Frequency Selectivity}
\begin{figure}[]
    \centering
    \includegraphics[width=0.8\linewidth]{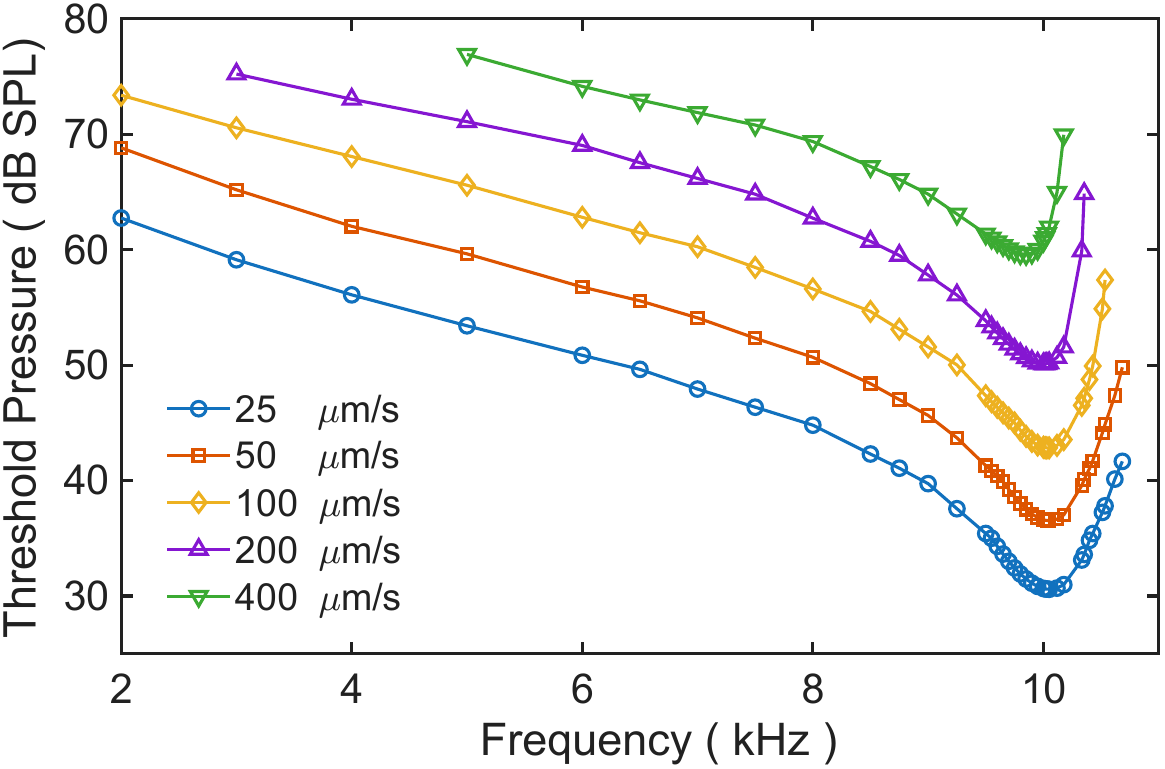}
    \caption{Model-predicted iso-response tuning curves. The lowest threshold occurs near the CF, demonstrating the greatest sensitivity of the system at this frequency.}
    \label{fig:TC}
\end{figure}
\begin{figure}[htbp]
\centering
\includegraphics[width=\linewidth]{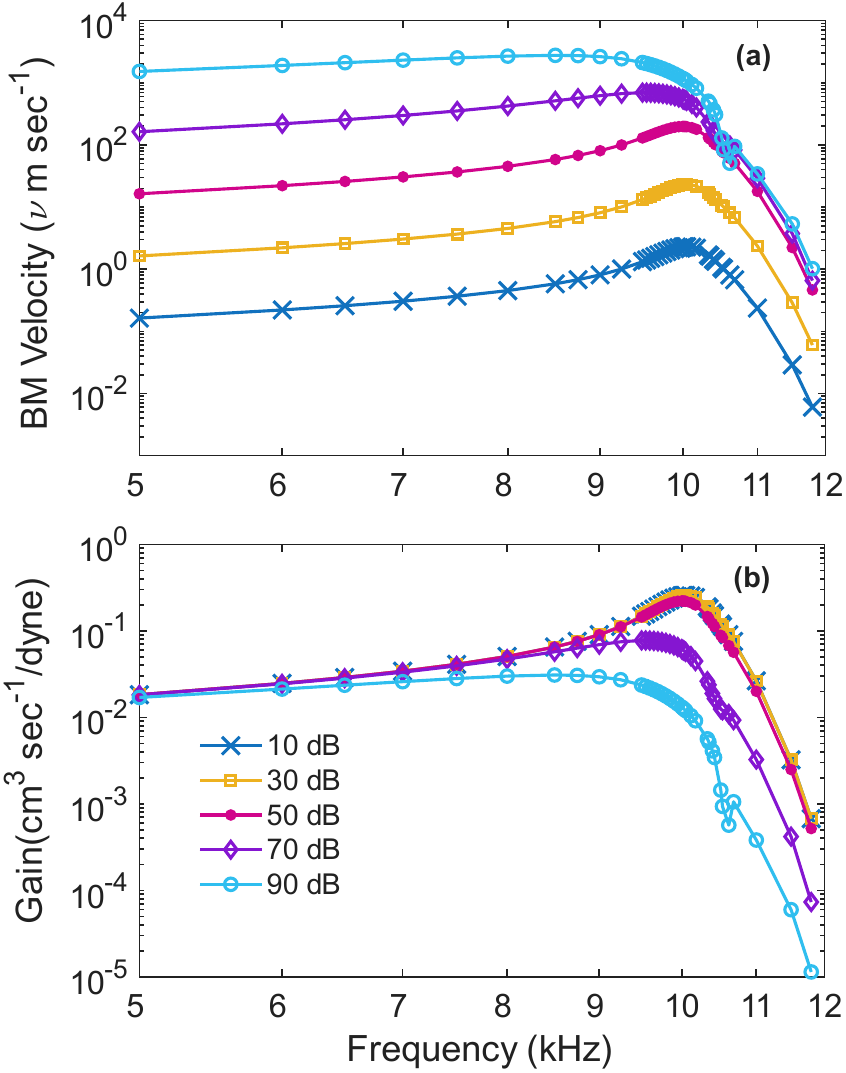}
\caption{Frequency selectivity characteristics of the model at stimulus levels of $10$, $30$, $50$, $70$ and $90$ dB SPL. 
(a) Velocity response and (b) the corresponding gain. The lowest stimulus levels produce the highest gain, and increasing the 
level progressively broadens the response, reducing frequency selectivity.
}
\label{fig:selectivity_panel}
\end{figure}

Frequency selectivity is investigated using iso-response tuning curves together with resonance and gain curves 
measured at fixed stimulus levels. The iso-response tuning curves describe the threshold stimulus required to produce 
a specified response, whereas the resonance and gain curves characterize the frequency response of the system at fixed stimulus levels.
Fig. \ref{fig:TC} shows the iso-response tuning curves obtained for target velocities ranging from $25$ ${\mu m s^{-1}}$ to $400$ $\mu ms^{-1}$.  
All curves show a minimum near the CF, indicating the greatest sensitivity of the system at this frequency.
Away from the CF, progressively larger stimuli are required to produce the same response magnitude. As the target response increases, 
the minima of the tuning curves shift marginally towards lower frequencies.
The frequency selectivity of the tuning curves is quantified using the quality factor $Q_{10}$, defined as the CF 
divided by the $10 - $ $\text{dB}$ bandwidth. A comparison between the experimentally measured and model-predicted $Q_{10}$ 
values is presented in Table \ref{tab:Q10}. The model gives $Q_{10}$ values comparable to the experimental values seen
\cite{ruggero1997basilar,rhode2007basilar}.

Fig. \ref{fig:selectivity_panel}(a)  shows the resonance curves obtained at different stimulus levels. 
At each stimulus level, the response exhibits a pronounced peak near the CF. As the stimulus level increases, 
the resonance peak shifts towards lower frequencies and becomes broader, indicating a reduction in frequency selectivity.  
Fig. \ref{fig:selectivity_panel}(b) shows the iso-intensity curves representing the gain. 
The lowest stimulus levels produce the highest peak gains. As the stimulus level increases, the peak gain decreases, 
and the frequency response becomes broader. The gain curves corresponding to $10$ to $50$ dB SPL exhibit comparable 
peak gains. At higher stimulus levels, the peak locations progressively separate, reflecting the level-dependent 
changes in cochlear amplification associated with the onset of nonlinear compression. Beyond the CF, the gain decreases rapidly.
\subsubsection*{Phase}
\begin{figure}[htbp]
    \centering
    \includegraphics[width=0.8\linewidth]{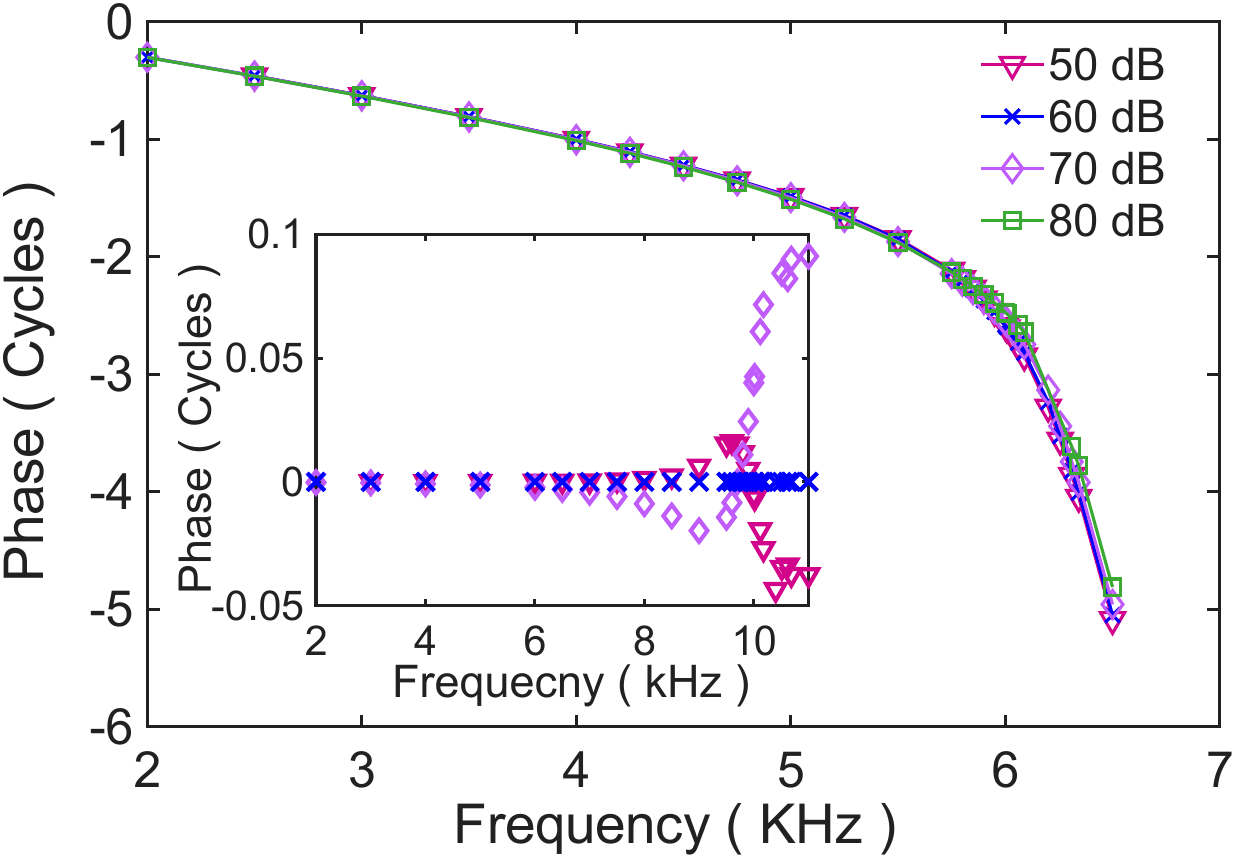}
    \caption{Phase accumulation as a function of stimulus frequency at different stimulus levels. 
    The inset shows the phase relative to the $60$ dB SPL response, highlighting the level-dependent phase accumulation near the CF.}
    \label{fig:Phase}
\end{figure}
Fig. \ref{fig:Phase} shows the phase accumulation as a function of the stimulus frequency for different stimulus levels. 
For frequencies below the CF, the phase varies gradually with frequency, resulting in a shallow slope that is largely 
independent of the stimulus level. As the CF is approached, the phase curves become considerably steeper, indicating a rapid 
accumulation of phase. Near the CF, the phase accumulation is approximately $2.5$ cycles. The inset in Fig. \ref{fig:Phase}
shows the response phase for $50$ and $70$ dB SPL stimulus, relative to the response for $60$ dB SPL stimulus. At frequencies 
below the CF, the response at lower stimulus levels leads that at higher stimulus levels. Above the CF, 
the lead-lag relationship reverses.
\begin{figure}[]
    \centering

    \includegraphics[width=0.75\linewidth]{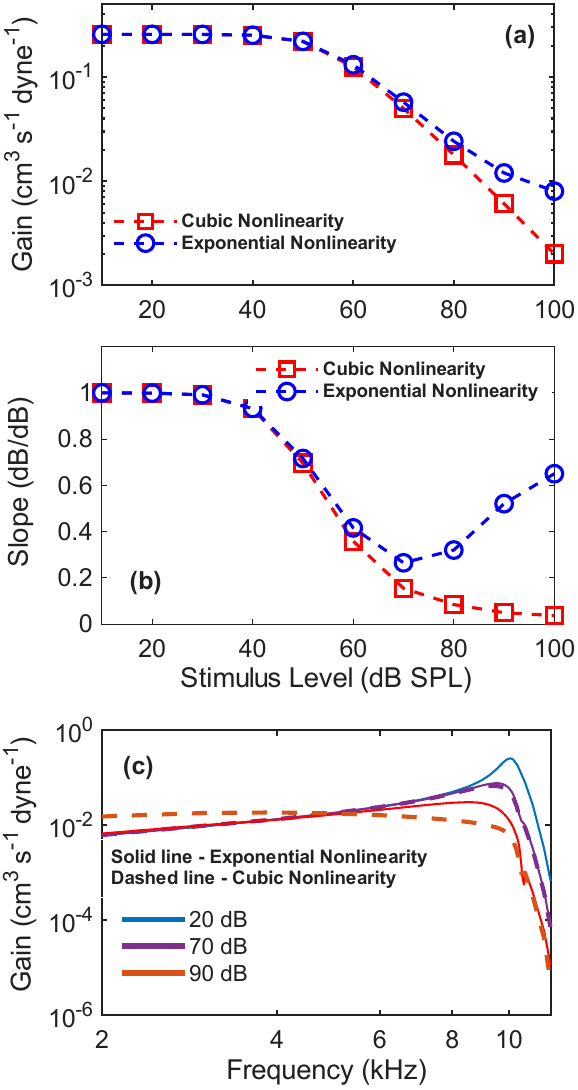}
   
\caption{Comparison between the exponential and cubic nonlinear models.
(a) Gain at the CF as a function of stimulus level.
(b) Local slope of the response as a function of stimulus level. 
(c) Frequency response curves at stimulus levels of $20$, $70$ and $90$ dB SPL. Both models exhibit nearly identical 
responses at low stimulus levels. At higher stimulus levels, the exponential model retains sharper frequency selectivity and exhibits a gradual convergence towards passive response. }
\label{fig:comparision_full_truncated_panel}
\end{figure}

\subsubsection*{Effect of Higher-Order Nonlinearities}

To investigate the role of higher-order nonlinearities, the responses obtained from the exponential nonlinear model are 
compared with those from a truncated approximation retaining up to cubic-order nonlinear terms. 
At low and moderate stimulus levels, up to $60$ $\text{dB SPL}$, both models produce nearly identical responses. 
As shown in Fig. \ref{fig:comparision_full_truncated_panel}, the gain at the CF, the local slope, and the 
iso-intensity gain curves are indistinguishable over this range. Beyond $60$ $\text{dB SPL}$, clear differences emerge between the two models. 
The exponential nonlinear model maintains a higher gain at the CF. 
The cubic nonlinear approximation exhibits a steeper reduction in gain with increasing stimulus level, as shown in \ref{fig:comparision_full_truncated_panel}(a). 
Fig. \ref{fig:comparision_full_truncated_panel}(b) shows that both models exhibit similar compression up to $60$ $\text{dB SPL}$. 
However, beyond this level, the behavior of the two models differs qualitatively. 
The exponential nonlinear model exhibits a recovery of the local slope at high stimulus levels. 
The cubic nonlinear approximation shows a stronger compression, with the local slope continuing to decrease throughout the stimulus range. 

A similar trend as \ref{fig:comparision_full_truncated_panel}(a) is reflected in the iso-intensity gain curves shown in \ref{fig:comparision_full_truncated_panel}(c). 
At low and moderate stimulus levels, the exponential nonlinear model and the cubic nonlinear approximation produce identical peak gains. 
At higher stimulus levels, the peak response of the cubic nonlinear approximation decreases more rapidly, accompanied by a progressive 
broadening of the gain curves. By $90$ $\text{dB SPL}$, the gain peak is no longer well defined.
The iso-response $Q_{10}$ values (Table \ref{tab:Q10}) obtained from the exponential nonlinear model and the cubic 
nonlinear approximation give similar values for the response levels considered. The small difference in $Q_{10}$ value for the
$400$ $\mu \text{m s}^{-1}$ iso-response curve is due to the fact that at this response level the stimulus amplitudes need to be large
and this is the regime where the two models deviate.
The traveling wave responses in Fig. \ref{fig:TW_panel}(a) and (b) further illustrate the differences between the two models. 
At $20$ $\text{dB SPL}$, the exponential nonlinear model and the cubic nonlinear approximation produce identical spatial displacement profiles. 
However, at $90$ $\text{dB SPL}$, the exponential nonlinear model produces a relatively more localized peak, whereas the cubic 
nonlinear approximation produces a broader peak with a greater shift towards the base.
These results demonstrate that the cubic nonlinear approximation provides an accurate description 
of the cochlear response for low and moderate stimulus levels. At higher stimulus levels, however, higher-order nonlinearities 
become essential for preserving the peak response, frequency selectivity, and the level-dependent growth of the response.

\subsubsection*{Effect of longitudinal coupling}
\begin{figure}[htbp]
\centering

\includegraphics[width=0.75\linewidth]{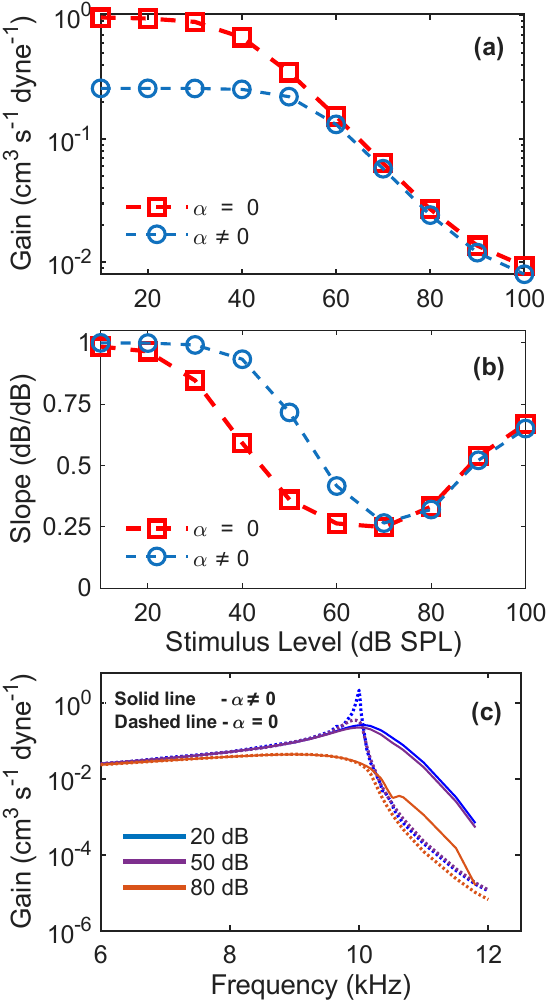}
\caption{Comparison of the exponential model with ($\alpha\neq0$) and without ($\alpha=0$) longitudinal coupling. 
(a) Gain at the CF as a function of stimulus level. (b) Local slope of the response as a function of stimulus level. 
(c) Gain curves at stimulus levels of $20$,$50$, and $80$ dB SPL. Longitudinal coupling broadens the frequency 
response and reduces the gain at low stimulus levels; its influence becomes progressively smaller with increasing stimulus level.}
\label{fig:comparision_coupling}
\end{figure}
\begin{figure}[htbp]
    \centering
    \includegraphics[width=\linewidth]{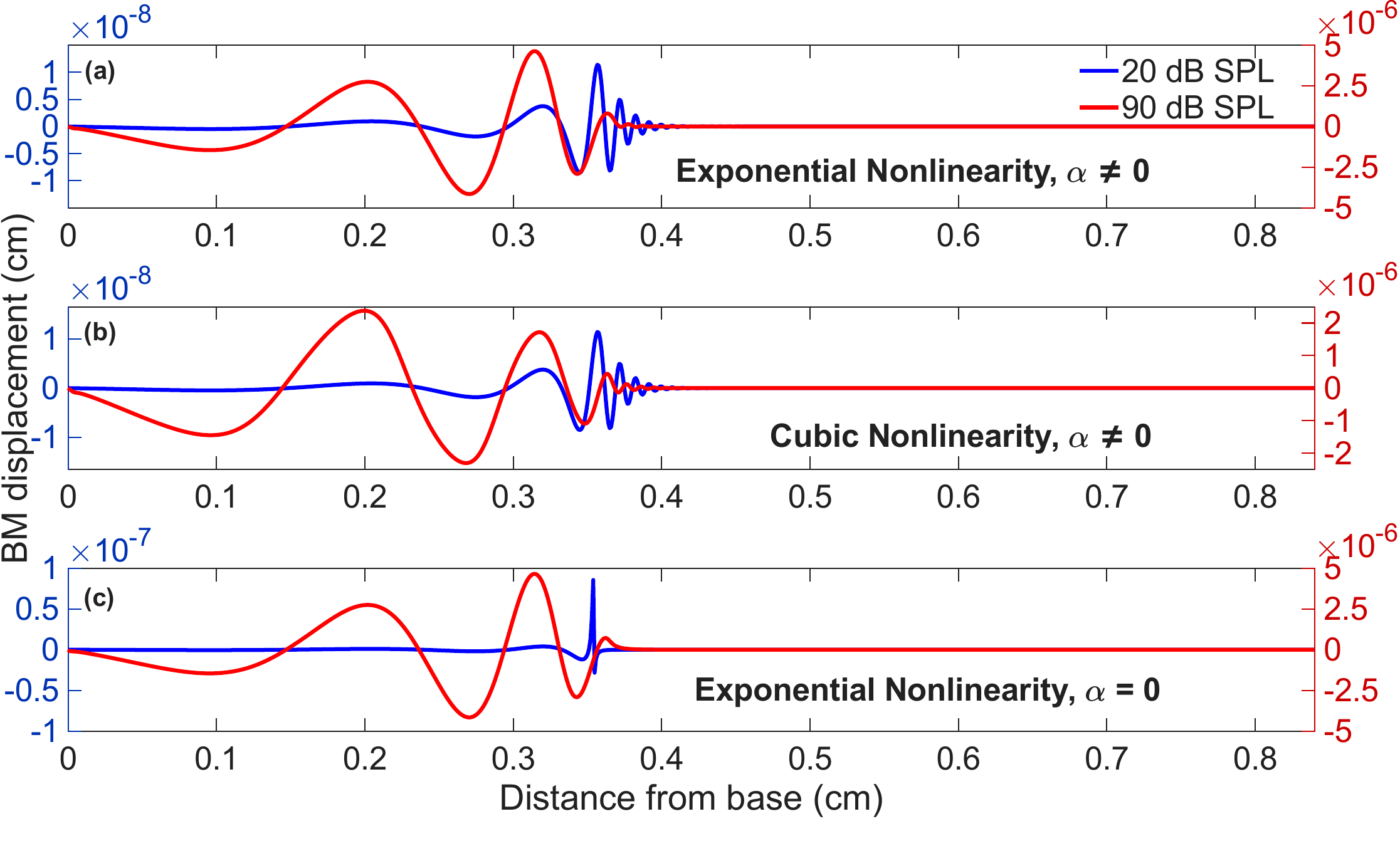}
\caption{Traveling-wave profiles at $20$ dB SPL and $90$ dB SPL for (a) The exponential model with longitudinal 
coupling ($\alpha\neq0$), (b) The cubic nonlinear approximation with longitudinal coupling, and (c) the 
exponential model without longitudinal coupling ($\alpha=0$). At $20$ dB SPL, the exponential and cubic models 
exhibit nearly identical traveling-wave profiles. At $90$ dB SPL, the cubic model exhibits a basal shift 
of the response maximum, whereas the exponential model retains a localized peak. At $20$ dB SPL, longitudinal 
coupling broadens the traveling-wave profile compared with the uncoupled model. At $90$ dB SPL, the coupled and 
uncoupled exponential models exhibit similar traveling-wave profiles. }
\label{fig:TW_panel}

\end{figure}

To examine the effect of longitudinal coupling, the responses obtained from the exponential nonlinear model 
are compared before and after the inclusion of longitudinal coupling.
Fig. \ref{fig:comparision_coupling}(a) compares the gain at the CF.
In the absence of longitudinal coupling, the gain remains substantially higher at low and moderate stimulus levels. 
The introduction of longitudinal coupling reduces the gain over the entire stimulus range. 
As the stimulus level increases, the differences in the gain of the two models become small.
The same trend is observed in the local slope shown in Fig. \ref{fig:comparision_coupling}(b). 
Without longitudinal coupling, the local slope begins to decrease at lower stimulus levels, implying that compression 
in the growth of response sets in early. The coupled model maintains a near to linear response at lower stimulus levels. 
Beyond $70$ $\text{dB SPL}$, the local slopes of the two models recover and become nearly identical. 

The iso-intensity gain curves are shown in Fig. \ref{fig:comparision_coupling}(c).  
At $20$ $\text{dB SPL}$, the uncoupled model exhibits a sharper resonance peak with a higher gain. 
The introduction of longitudinal coupling reduces the peak gain and broadens the frequency response. 
At  $80$ $\text{dB SPL}$, the peak gains of the two models become comparable. However, the coupled model continues to exhibit a broader response, beyond the CF. 
The corresponding iso-response $Q_{10}$ values are listed in Table \ref{tab:Q10}. 
In the absence of longitudinal coupling, the tuning curves are sharper as reflected by the substantially larger 
$Q_{10}$ values. The introduction of longitudinal coupling broadens the tuning curves, resulting in reduced $Q_{10}$ 
values over the range of target response amplitudes.

The traveling wave responses in Fig. \ref{fig:TW_panel}(a) and (c) further illustrate the effect of longitudinal coupling. 
At $20$ $\text{dB SPL}$, the uncoupled model exhibits a highly localized traveling wave with a larger peak amplitude, 
whereas longitudinal coupling produces a broader traveling wave extending over a larger region of the basilar membrane.  
At $90$ $\text{dB SPL}$, the peak amplitudes of the two models are nearly identical. However, the coupled model exhibits a 
traveling wave with a more extended oscillatory tail. The influence of longitudinal coupling is most pronounced at low and 
moderate stimulus levels. At higher stimulus levels, the differences between coupled and uncoupled responses become small.

\begin{table}[h]
\centering
\caption{$Q_{10}$ values obtained from the iso-response tuning curves at different target velocities.}
\label{tab:Q10}
\begin{tabular}{ccccc}
\hline
\textbf{Target Velocity} &
\textbf{Experiment} &
\textbf{M1} &
\textbf{M2} &
\textbf{M3} \\
($\mu$m s$^{-1}$) & & & & \\
\hline
25  & 5.3 & 5.6 & 5.6 & 21 \\
50  & 5.6 & 5.7 & 5.7 & 51 \\
100 & 6.1 & 6.0 & 6.0 & 62 \\
200 & 7.1 & 5.9 & 5.9 & 23 \\
400 & 5.2 & 4.4 & 4.1 & 6.5 \\
\hline
\end{tabular}

\vspace{1mm}
\parbox{0.9\linewidth}{\footnotesize
\textit{Note:} M1: Exponential model with longitudinal coupling; M2: Cubic model with longitudinal coupling; M3: Exponential model without longitudinal coupling.
}
\end{table}

\section*{Discussion}
The objective of the present study was to model the cochlea using a minimal set of physically motivated ingredients. 
The model was used to investigate how the characteristic signatures of the cochlear active process emerge from such a description. 
Particular emphasis was placed on the role of the displacement-dependent active process in producing amplification, compressive nonlinearity, and frequency selectivity. 
The influence of fluid-mediated and nearest-neighbor longitudinal coupling on the responses of the BM-OC system was also examined. The contribution of higher-order nonlinearities was investigated by comparing the exponential nonlinear term with its cubic approximation.
\par
The coupled oscillator model qualitatively reproduces several characteristic features of cochlear mechanics. These include traveling wave propagation, amplification, compressive response growth, frequency selectivity, level-dependent phase behavior, and the experimentally observed trends in $Q_{10}$. Comparisons with the cubic approximation show that higher-order nonlinearities become important at large response amplitudes. Comparisons between the coupled and uncoupled models demonstrate the role of longitudinal coupling in determining traveling wave propagation, phase accumulation, and frequency selectivity.
\par
A central feature of the present model is the displacement-dependent nonlinear term.
This term was introduced to mimic the net active mechanical feedback provided by outer hair cells(OHCs), one of the sensory receptor cells  of the organ of Corti. 
In the mammalian cochlea, OHCs are believed to function as both sensors and mechanical effectors, supplying energy locally to the traveling wave to compensate for viscous dissipation.
This active feedback enhances traveling-wave amplification, frequency selectivity, and brings in compressive nonlinearity\cite{dallos1992active}. 
In the present model, this behavior is represented through the amplitude-dependent nonlinear damping term. 
At small basilar membrane displacements, the active contribution largely compensates for viscous damping, leading to amplification of weak responses. 
As the displacement increases, the active contribution is progressively reduced. 
Consequently, the response becomes compressive while maintaining sharp frequency selectivity.
\subsubsection*{Traveling-wave propagation}
An important aspect of cochlear mechanics is the spatial evolution of basilar membrane motion in response to acoustic stimulation.
The model successfully reproduces the characteristic traveling wave response of the cochlea, with the wave propagating from the basal end towards the apical end of the oscillator chain.
The evolution of the traveling wave profile with the stimulus level can be understood from the  Fig. \ref{fig:TW_panel}(a).
The active process strongly amplifies the response near the characteristic place by compensating for the dissipative damping. 
As a result, the traveling wave is strongly localized around the CF location. With the increase in stimulus level, 
two key changes occur: the location of the maximum response shifts towards the base and the spatial response becomes broader. 
 
\subsubsection*{Compressive Nonlinearity}
Further insight into the nonlinear behavior of the model can be understood from the level-dependent characteristics of the 
local response represented in Fig. \ref{fig:compression_panel}. 
The compressed growth of the response enables the ear to process a vast dynamic range of sound intensities while maintaining high sensitivity to weak stimuli. 
In the present model, this behavior is captured by the amplitude-dependent active term with $\beta e^{-\left(\frac{\eta(x,t)}{\eta_{\rm th}}\right)^2}$. Within this term, $\eta_{\rm th}$ defines the displacement scale beyond which active feedback is gradually attenuated.

The local rate of growth in Fig. \ref{fig:compression_panel}(b) provides a clear measure of the transition between different operating regimes. 
At low stimulus levels, the slope remains close to unity, indicating that the response remains in the small displacement regime. 
In this regime the active contribution nearly balances viscous damping. As the response amplitude increases, the compensation 
of the viscous damping is progressively reduced. This is reflected in the reduction in the local slope and emergence of compression. 
The minimum in the local slope corresponds to the region of the strongest compression. 
At a higher stimulus level, the local slope gradually increases again, suggesting that this mechanism is increasingly governed 
by the passive mechanical dynamics of the oscillator. In this regime, the system behaves like a coupled forced damped oscillator, 
resulting in a partial recovery towards linear response growth.

The simultaneous reduction in gain and preservation of relatively sharp tuning reflects a compromise between sensitivity and selectivity. 
The behavior of the gain-bandwidth product (Fig. \ref{fig:compression_panel}(d)) demonstrates that the reduction in 
amplification is not accompanied by a proportional loss of frequency selectivity. From a physiological perspective, this 
represents an important characteristic of cochlear mechanics. As stimulus levels increase, the active process becomes
less effective, preventing excessive amplification of intense sounds. The model naturally reproduces this feature through 
the progressive reduction of active feedback with the stimulus level.

\subsubsection*{Frequency selectivity}
 Compressive nonlinearity and sharp frequency selectivity are the two defining manifestations of the cochlear amplifier. 
Frequency selectivity enables the cochlea to resolve the spectral components of complex acoustic stimuli by producing a localized response near the CF. The preservation of frequency selectivity over a wide range of stimulus levels is therefore another essential function of the cochlear amplifier. 
\par
In the present model, both arise from the same amplitude-dependent active contribution. The iso-response tuning curves in Fig. \ref{fig:TC} and iso-intensity resonance curves in Fig. \ref{fig:selectivity_panel} provide descriptions of the same underlying frequency-selective behavior. 
At low stimulus levels, the resonance curves exhibit narrow peaks together with the highest gain near the CF, indicating a strong amplification and sharp frequency tuning. The gain changes only modestly over low and intermediate stimulus levels.
This suggests  that the cochlear amplifier is able to preserve both  amplification and frequency selectivity even as active contribution begins to reduce. At higher stimulus levels, the resonance curves broaden and their peaks shifts towards lower frequencies, accompanied by a reduction in gain. 

The spatial evolution of the traveling wave in Fig. \ref{fig:TW_panel} provides additional insight  on the reduction in frequency selectivity. 
As discussed earlier, low stimulus levels produce a traveling wave that is strongly localized near the characteristic place.
At higher stimulus levels, the response extends over a broader region of the basilar membrane.
Although the traveling wave profiles and resonance curves describe different aspects of cochlear response, they convey a consistent picture. 
The traveling wave profiles characterize the spatial extent of the basilar membrane excitation for a given stimulus frequency, 
whereas the resonance curves describe the frequency response of a single location. At higher stimulus levels, the responses 
become less localized in both the spatial and spectral domains, indicating a progressive reduction in cochlear selectivity. 
This behavior is consistent with the level-dependent contribution of the active nonlinear term. As the active 
contribution is reduced, the cochlear amplifier becomes less effective at maintaining sharp frequency tuning.
Consequently, individual locations respond over a broader range of frequencies, giving rise to broader resonance curves. 
Likewise, a given stimulus excites a broader region of the basilar membrane, resulting in a broader traveling wave envelope.
\par 
Taken together, these results show that the level-dependent active contribution governs both cochlear amplification and frequency selectivity. 
Its influence is evident in both the spatial and spectral organization of the cochlear response.

\subsubsection*{Role of higher-order nonlinearities}
The results presented so far demonstrate that the proposed active nonlinearity reproduces the principal features of cochlear mechanics across a wide dynamic range. To examine how the form of the active feedback influences these responses, we compare the exponential feedback with its weak-amplitude approximation. This approximation corresponds to the familiar cubic nonlinearity employed in many Van der Pol and Rayleigh-type cochlear oscillator models.
In the simplified model, the active feedback is represented by a cubic nonlinearity.
This comparison highlights the features of the cochlear response that are preserved in the simplified model, as well as those that differ from the exponential nonlinear model.\par
At low and moderate stimulus levels, both forms of nonlinearities produce nearly identical responses. The gain at the CF, the local response slope, the resonance curves in Fig. \ref{fig:comparision_full_truncated_panel}(c) and the traveling wave profiles in Fig. \ref{fig:TW_panel}(a) and(b) are in close agreement, indicating that the simplified model captures the essential features of the cochlear response within this operating regime.\par
At higher stimulus levels, the two forms of nonlinearities exhibit qualitatively different behavior. 
The exponential nonlinear model preserves a pronounced resonance peak together with a larger gain at the CF. 
The cubic approximation predicts a more rapid reduction in gain and a progressive degradation of the resonance peak. 
The local response slope also evolves differently in the two models. 
While the exponential model exhibits a partial recovery of the local slope at high stimulus levels, the cubic approximation continues to predict a monotonic decrease throughout the stimulus range. 
This difference reflects the distinct manner in which the two nonlinearities regulate the active process as the response amplitude increases. \par 
The traveling wave responses provide a complementary perspective on these differences. At low stimulus levels, both models produce identical traveling wave profiles. 
At high stimulus levels, however, the maximum response predicted by the cubic approximation is displaced further towards the base than in the exponential model.
Consequently, by the time the traveling wave reaches the characteristic place, its amplitude has already decreased substantially. 
This behavior is consistent with the reduced gain at the CF and the degradation of the resonance peak predicted by the cubic approximation. 
The continued reduction in the local response slope should therefore be interpreted together with the traveling wave response, since the response measured at the characteristic place is influenced not only by the local nonlinear dynamics but also by the spatial redistribution of the traveling wave maximum. \par
Taken together, these results show that the form of the active nonlinearity plays a central role in reproducing the cochlear mechanics over a wide dynamic range. The present comparison is also consistent with the development of Van der Pol and Rayleigh-type cochlear oscillator models. The quadratic dependence of the active feedback underlying the familiar cubic nonlinearity accurately captures the weakly nonlinear regime. Previous studies improved the high-level response by introducing progressively more mathematically involved damping profiles. In contrast, the exponential active feedback employed here naturally contains the higher-order contributions required at larger response amplitudes, allowing the transition from active amplification to passive behavior to emerge within a single physically motivated feedback law.

\subsubsection*{Effect of longitudinal coupling} 
While the local response of each oscillator is governed by the balance between passive damping and active feedback, the cochlear response also depends on the mechanical interaction between neighboring regions of the BM-OC system. Together, they distribute the response over a finite region of the basilar membrane, broadening the spatial extent of the travelling wave.
In the present model, these interactions are represented by elastic and dissipative longitudinal coupling between the adjacent elements. 

The role of longitudinal coupling is illustrated through the characteristic-frequency gain, gain curves, and local response slopes (Fig. \ref{fig:comparision_coupling}), together with the traveling-wave profiles (Fig. \ref{fig:TW_panel}(a) and (c)).
\par 

The most immediate consequence of longitudinal coupling is observed in the frequency response. 
Compared with the uncoupled model, the coupled system exhibits a lower gain at the CF together with a broader resonance peak. 
This broadening is reflected in the reduced $Q_{10}$ values in Table \ref{tab:Q10}
over the range of target response amplitudes, indicating a modest reduction in the sharpness of tuning. 
Rather than concentrating the response at a single location, longitudinal coupling causes neighboring oscillators to participate in the response, producing a broader frequency tuning that is more consistent with the distributed nature of cochlear mechanics.
\par
Longitudinal coupling also influences the level dependence of the response. 
At low and intermediate stimulus levels, the coupled model maintains a nearly linear local response over low stimulus levels, delaying the onset of compressive nonlinearity relative to the uncoupled model. 
However, with the increase in the stimulus level, the differences in both the local response and the CF gain become progressively smaller. 
These observations indicate that the influence of longitudinal coupling is most pronounced at low and intermediate stimulus levels.\par
The traveling wave response provides a clear illustration of the role of longitudinal coupling. 
At low stimulus levels, coupling broadens the traveling wave envelope. This allows the response to spread over a larger region of the basilar membrane while reducing the peak amplitude at the CF location.
At higher stimulus levels, peak amplitudes in the coupled and the uncoupled models become comparable. 
However, the coupled model still exhibits a more extended oscillatory tail. These observations demonstrate that longitudinal coupling influences not only the magnitude of the local response but also the spatial extent over which the response is distributed. \par
These results show that the longitudinal coupling complements the local active process by governing the collective response of neighboring oscillators. 
The balance between passive damping and active feedback determines the local amplification and compressive behavior. 

\subsubsection*{Phase}
The traveling-wave envelope and phase response provide complementary descriptions of cochlear wave propagation.
While the traveling wave envelope describes the spatial distribution of the response, the phase response characterizes its propagation along the cochlea( Fig. \ref{fig:Phase} and \ref{fig:TW_panel}). 
The phase response predicted by the model shows a gradual variation below the CF, followed by a rapid accumulation as the CF is approached. This qualitative feature is consistent with the experimentally observed behavior. The model predicts a phase accumulation of $2.6$ cycles at the CF, in close agreement with the experimentally observed value of about $2.5$ cycles \cite{ruggero1997basilar}. The experimentally observed level-dependent phase lead-lag reversal about the CF is also reproduced by the model.  The exponential and cubic nonlinearities exhibit nearly identical phase responses at low stimulus levels, consistent with their similar traveling wave profiles. At higher stimulus levels, the cubic nonlinearity predicts a modest increase in the accumulated phase, although the overall phase behavior remains similar. 
The introduction of longitudinal coupling increases the accumulated phase, consistent with the broader traveling wave envelope observed in the coupled model. 
These results indicate that the model captures the essential wave-propagation characteristics of cochlear mechanics.

\section*{Summary}
In this work, we introduce a physically motivated critical-oscillator model for the mammalian cochlea. 
Unlike most existing critical-oscillator approaches, which are based on the normal form of the Hopf bifurcation, the current model is derived from a more physically grounded dynamical system. Although the system undergoes a 
Hopf bifurcation and operates in its vicinity, its response at large input amplitudes differs significantly 
from that of conventional normal form-based models. The proposed formulation also shares the underlying philosophy of Van der Pol and Rayleigh-type cochlear oscillator models, in which active feedback is introduced through nonlinear damping. By combining operation near a Hopf bifurcation with a physically grounded nonlinear damping mechanism, the present model provides a unified framework for describing both the weakly nonlinear and the strongly nonlinear regimes of cochlear mechanics \cite{duifhuis2011hopf}.In the weak-response regime, the exponential feedback naturally recovers the familiar cubic nonlinearity. At larger response amplitudes, it retains the higher-order nonlinearities required to describe the response over a wide dynamic range. This allows the transition from active amplification to predominantly passive behavior to emerge naturally, without introducing additional phenomenological nonlinear feedback laws. Therefore, the proposed formulation belongs to the broader family of nonlinear cochlear oscillator models while preserving a compact physical interpretation. More generally, it satisfies three important modeling principles \cite{diependaal1983department}. First, the active and nonlinear processes are intrinsically linked through a common feedback mechanism. Second, passive cochlear mechanics are recovered when the active process is removed. Finally, active amplification remains spatially localized along the basilar membrane. Spatial interactions constitute another ingredient of cochlear mechanics.
Previous studies have emphasized the role coupling between the active elements \cite{ver2024marrying,kern2003essential}.
In the present model, these interactions are represented through fluid-mediated coupling and nearest-neighbour longitudinal coupling.

We show that even this minimal model qualitatively reproduces several key characteristics of cochlear 
mechanics. In particular, it captures the observed phase behavior, realistic $Q_{10}$ values, 
the gain curve at the CF, and the gain–bandwidth relationship. 
Our analysis further demonstrates that both oscillator coupling and higher-order nonlinearities 
are essential for reproducing experimental observations. Specifically, the low-amplitude behavior 
of the gain curve depends critically on the coupling between neighboring oscillators, whereas 
the experimentally observed local slope behavior of the input-output curve requires the 
inclusion of higher-order nonlinear terms. These results indicate the importance of incorporating 
both coupling and nonlinear effects beyond the Hopf normal form in developing realistic cochlear models.\par
While the proposed model successfully reproduces several qualitative features of cochlear mechanics, 
there remain important discrepancies between its predictions and experimental observations. We list some
of them here:
(i) The model does not capture the nearly linear and substantial response at the CF location to 
high-frequency stimulus at large stimulus amplitudes. In measurements of the basilar membrane response, 
significant activity is often present at frequencies above the CF for strong inputs, 
whereas the present model predicts a more pronounced attenuation in this regime.
(ii) The response predicted by the model at the CF location on the low-frequency side of the 
CF decays more slowly than is observed experimentally. 
Cochlear measurements typically show a sharper drop toward lower frequencies, indicating stronger 
frequency selectivity than that produced by the current model. This discrepancy suggests that additional 
mechanisms influencing wave propagation and/or energy dissipation may need to be incorporated.
(iii) The predicted response at the CF location to a stimulus at low amplitude with a CF 
is approximately five times smaller than that measured experimentally. Although the model reproduces 
the qualitative form of the gain curve, it underestimates the amplification achieved by the biological 
cochlea in the weak-signal regime. This indicates that the active amplification mechanisms represented 
in the model are not yet sufficient to explain the observed sensitivity.

We now briefly discuss possible extensions/modifications of the model that could lead to better quantitative agreement 
between the model and experimental data. In the current model, the active force is in phase with the velocity of the
membrane segment: (i) Any feedback mechanism that generates an active process will, in general, have a delay. 
One can generalize the current model to introduce a phase lag between the active force and velocity to account for
this delay.  (ii) One could introduce a velocity-dependent cutoff mechanism rather than one
that is based on the displacement cutoff. Such a modification could provide an alternative representation of 
the nonlinear active processes within the cochlea and may help improve the model's behavior at both high 
frequencies and large stimulus amplitudes. (iii) Another important direction is to consider a two-dimensional 
description of cochlear dynamics. The present model is effectively one-dimensional and therefore neglects 
transverse variations that may influence the behavior. The one-dimensional approximation
requires that the wavelength of the membrane excitation is much less than the height of the cochlear chamber.
This approximation fails for various input conditions. A two-dimensional framework could provide a more 
faithful representation of cochlear structure.\par
Further improvements may also be achieved through systematic parameter optimization. In the current work, 
several parameters are chosen based on general physical considerations rather than detailed experimental constraints. 
Introducing spatially varying properties such as mass density of the membrane, coupling strength, 
damping, and nonlinear coefficients, guided either by direct experimental measurements or by data-driven fitting procedures, could substantially enhance the quantitative accuracy of the model. Taken together, these 
extensions provide a clear path toward constructing a more realistic critical oscillator 
based cochlear model capable of reproducing both the qualitative and quantitative features of 
experimentally observed cochlear responses.

\newpage
\printbibliography
\end{document}